\documentclass{llncs}  % do not change this line!

\usepackage{booktabs}
\usepackage{adjustbox}
\usepackage{varwidth}
\usepackage{parskip}
\usepackage{amssymb,amsmath}

\usepackage[colorlinks=true,urlcolor=blue,linkcolor=blue,citecolor=blue]{hyperref}

\usepackage{tikz}
\usetikzlibrary{arrows,shapes, shapes.geometric, decorations,automata,backgrounds, shapes}
\tikzset{elliptic state/.style={draw,ellipse}}
\usepackage{float}
\usepackage{enumerate}
\begin{document}

\title{Modeling Normative Multi-Agent Systems from a Kelsenian Perspective}  % put your title here!
%\titlenote{Produces the permission block, and copyright information}

% AAMAS: as appropriate, uncomment one subtitle line; check the CFP
%\subtitle{Extended Abstract}
%\subtitle{Industrial Applications Track}
%\subtitle{Socially Interactive Agents Track}
%\subtitle{Blue Sky Ideas Track}
%\subtitle{Robotics Track}
%\subtitle{JAAMAS Track}
%\subtitle{Doctoral Mentoring Program}

%\subtitlenote{The full version of the author's guide is available as \texttt{acmart.pdf} document}

% AAMAS: submissions are anonymous for most tracks
%\author{Paper \#XXX}  % put your paper number here!

%% example of author block for camera ready version of accepted papers: don't use for anonymous submissions
%
\author{Christiano Braga\inst{1}, E. Hermann Haeulser\inst{2} and J\'essica S. Santos\inst{1}}
\institute{Instituto de Computa\c{c}\~ao, Universidade Federal Fluminense, }
%\authornote{Dr.~Trovato insisted his name be first.}
%\affiliation{
  \institute{Instituto de Computa\c{c}\~ao, Universidade Federal Fluminense \and Departamento de Inform\'atica, Pontif\'icia Universidade Cat\'olica do Rio de Janeiro}

\maketitle

\begin{abstract}  % put your abstract here!
Standard Deontic Logic (SDL) has been used as the underlying logic to model and reason over Multi-Agent Systems governed by norms (NorMAS). It is known that SDL is not able to represent contrary-to-duty (CTD) scenarios in a consistent way. That is the case, for example, of the so-called Chisholm paradox, which models a situation in which a conditional obligation that specifies what must be done when a primary obligation is violated holds. In SDL, the set of sentences that represent the Chisholm paradox derives inconsistent sentences. Due to the \emph{autonomy} of the software agents of a NorMAS, norms may be violated and the underlying logic used to model the NorMAS should be able to represent \emph{violation scenarios}. The contribution of this paper is threefold: (i) we present how Kelsenian thinking, from his jurisprudence in the context of legal ontologies, and Intuitionist Hybrid Logic can be adopted in the modeling of NorMAS, (ii) discuss how this approach overcomes limitations of the SDL and (iii) present a discussion about normative conflict identification according to Hill's functional taxonomy, that generalizes from standard identification by impossibility-of-joint-compliance test.
\iffalse
A multi-agent system (MAS) is a collection of autonomous entities, called agents, that interact to fulfil the system’s intended behavior. When regulated by a collection of rules, also called norms, such a system is referred as a Normative MAS (NorMAS). The standard technique to reason on NorMAS uses Deontic Logic as underlying logic where norms becomes formulae in the associated theory. However, it is known to fail on the modeling of contrary-to-duty scenarios, also called deontic paradoxes, since the resulting theories turn out to be inconsistent. In this paper, we propose a new approach to reason on NorMAS, based on Intuitionistic Hybrid Logic (IHL) and Kelsenian Jurisprudence. Essentially, norms are represented as nominals in the associated IHL theory. We discuss normative conflict identification according to Hill’s functional taxonomy, that generalizes from standard identification by impossibility-of-joint- =compliance test. Norm conflicts are resolved by norm precedence, naturally captured by the underlying Heyting algebra.
\fi
\end{abstract}

% AAMAS: the ACM CCS are not needed within AAMAS papers
%%
%% The code below should be generated by the tool at
%% http://dl.acm.org/ccs.cfm
%\begin{CCSXML}
%<ccs2012>
%  <concept_id>10010520.10010553.10010562</concept_id>
%  <concept_desc>Computer systems organization~Embedded systems</concept_desc>
%  <concept_significance>500</concept_significance>
%  <concept_id>10010520.10010575.10010755</concept_id>
%  <concept_desc>Computer systems organization~Redundancy</concept_desc>
%  <concept_significance>300</concept_significance>
%  <concept_id>10010520.10010553.10010554</concept_id>
%  <concept_desc>Computer systems organization~Robotics</concept_desc>
%  <concept_significance>100</concept_significance>
%  <concept_id>10003033.10003083.10003095</concept_id>
%  <concept_desc>Networks~Network reliability</concept_desc>
%  <concept_significance>100</concept_significance>
%</ccs2012>  
%\end{CCSXML}
%
%\ccsdesc[500]{Computer systems organization~Embedded systems}

  % put your semicolon-separated keywords here!

\section{Introduction}
\label{sec:intro}
Standard Deontic Logic (SDL) is the traditional logic for reasoning about normative aspects, such as, obligations, permissions and prohibitions \cite{vonWright1951}. This logic has been widely applied to formalize Normative Multi-Agent Systems (NorMAS) \cite{broersen2013}, which are Multi-Agent Systems (MAS) that adopt norms as a means to coordinate and restrict the behavior of software agents, aiming at achieving the overall goal of the system. In NorMAS, norms can be viewed as a regulatory mechanism that guides software agents, by establishing the desirable/ideal behavior of them.

It is known that there are scenarios that only can be represented in SDL by sets of sentences that derive inconsistent sentences or derive sentences with a counterintuitive reading \cite{MeyerDignumWieringa1994}. Such scenarios are called contrary-to-duty (CTD) paradoxes. The CTD paradoxes that are counterintuitive in the common sense reading are not considered so critical as the CTD paradoxes that derive inconsistent sentences \cite{mcnamara2006}. The Chisholm paradox \cite{chisholm1963} is a scenario that results in an inconsistent theory when formalized in SDL. This paradox models a situation in which: (1) there is a primary obligation stating what ought to be done; (2) there is a compatible-with-duty obligation that says what must be done when the primary obligation is fulfilled; (3) there is a conditional obligation (also called CTD obligation) that determines what must be done if the primary obligation is violated; and (4) there is a factual claim that indicates that the primary obligation was violated. Note that a CTD obligation is an obligation that is only enforced when a violation occurs.

The SDL formalization of the Chisholm paradox is the conjunction of the following sentences:

\begin{eqnarray}
\label{eqn:op} O(p) \\
\label{eqn:opq}O(p \Rightarrow q) \\
\label{eqn:nponq}\neg p \Rightarrow O(\neg q) \\
\label{eqn:np}\neg p
\end{eqnarray}
where $O$ is the deontic modality for obligation, $p$ is a propositional variable and so is $q$. 

SDL has the following axioms and inference rules:
\[\begin{array}{rl}
\mathbf{TAUT} & \mbox{All the tautologies of classical propositional logic.} \\
\mathbf{K} & O(p \Rightarrow q) \Rightarrow (O(p) \Rightarrow O(q)) \\
\mathbf{D} & O(p) \Rightarrow \neg O(\neg p) \\
\mathbf{MP} & \mbox{if } \vdash p \mbox{ and } \vdash p \Rightarrow q \mbox{ then } \vdash q \\
\mathbf{OB\mbox{-}NEC} & \mbox{if } \vdash p \mbox{ then } \vdash O (p).
\end{array}\]

As for the proof of inconsistency, from (\ref{eqn:opq}) and $\mathbf{K}$ we get $O(p) \Rightarrow O(q)$, and then from (\ref{eqn:op}) and $\mathbf{MP}$, we get $O(q)$, but by $\mathbf{MP}$ alone we get $O(\neg q)$ from (\ref{eqn:nponq}) and (\ref{eqn:np}). From these two conclusions, by Propositional Calculus, we get $O(q) \land O(\neg q)$, contradicting the SDL principle that obligations cannot conflict~\cite{vonWright1951}, formally $\vdash \neg(O(q) \land O(\neg q))$, i.e., $q$ cannot be obligatory and forbidden simultaneously.

The impossibility of representing CTD scenarios is a relevant limitation of SDL since in many applications norms can be violated. This kind of reasoning is needed in NorMAS because software agents are endowed with autonomy and sometimes the actual behavior of an agent deviates from the ideal behavior (according to the norms), i.e., an agent may choose not to comply with the system's norms in order to achieve its own goals \cite{broersen2013,carmo2002}. Then, it is natural to define CTD obligations in MAS telling what should be done when a norm is violated. Applications that deal with fault-tolerance, for instance, also need to represent CTD scenarios, i.e., once an obligation is violated, a corrective action must be performed \cite{castro2009}.

Another challenge, together with the modeling of CTD scenarios, that must be taken into account in NorMAS, is that the system must be able to deal with conflicts among norms. Two norms are in conflict, for instance, when one obliges an agent to perform an action that is being prohibited by other norm simultaneously. There is a famous example in the literature illustrating this situation, where a soldier is obliged to kill but he has a (moral) obligation to not to kill. When there is a normative conflict, the agent cannot choose to comply with both norms and whatever the agent does or refrain from doing will lead to a norm violation \cite{vasconcelos2009}. In MAS, considering that an agent can play different roles, it is common that conflicts of this nature arise between norms associated with the different roles~\cite{cholvy1995}. (Several techniques to deal with normative conflicts are nicely summarized in~\cite{santos2017}.)

In this paper we propose a novel approach to formalize NorMAS that overcomes limitations of SDL. We represent norms from a \emph{Kelsenian perspective} with Intuitionistic Hybrid Logic (IHL)~\cite{brauner2006} as underlying logic. The Kelsenian perspective does not assign truth-values to norms. Rather, they are understood as \emph{worlds} (in a Kripke structure) where properties of a MAS hold or not. In Hybrid Logic terminology, norms are understood as nominals.    
The second contributions of this paper is a discussion on how NorMAS with a Kelsenian interpretation of norms, or Kelsenian NorMAS for short, deal with normative conflicts while considering Hill's taxonomy~\cite{hill}. 

The remainder of this paper is organized as follows. In Section~\ref{sec:related} we present some approaches that also do not consider SDL as a suitable way to represent normative systems. Section~\ref{sec:knormas} presents the concept of norm according to the Kelsenian Jurisprudence and shows how Intuitionistic Hybrid Logic (IHL) complies with Kelsenian requirements. In Section~\ref{sec:CTDKnormas}, we present an example of CTD scenario in a NorMAS and discuss how it is modeled in our approach. Section~\ref{sec:hill-knormas} describes how our approach represents normative conflicts taking into account the Hill's taxonomy for normative conflicts. Section~\ref{sec:conclusion} concludes this paper and presents directions for future work.

\section{Related Work}
\label{sec:related}

%There are many approaches to model NorMAS, and to detect and solve normative conflicts, nicely summarized in~\cite{santos2017} but there are still many open issues~\cite{gabbay}. One open issue pointed out in ~\cite{gabbay}

The approach described in~\cite{castro2009deontic} describes a deontic action logic with stratified norms for designing and reasoning about fault-tolerant systems. When a violation occurs in these systems, a non-desirable system state is reached and CTD norms are applied to recover the system. In order to be able to represent CTD norms, the logic described considers that there are different levels of norms in the specification, and violations of norms are tolerated only at some levels.

In~\cite{haeusler2011}, the third author and others propose Intuitionistic Description Logic (iALC), for reasoning about laws. iALC is a notation variant of IHL, designed to cope with the ontological requirements in~\cite{haeusler2010}. In particular, iALC handles Commitment~\ref{it:law-conn}, discussed in Section~\ref{sec:knormas}, very nicely as iALC concepts represent different normative systems (understood as a collection of norms) quite naturally.
%IHL, has support for normative systems (by the partial order on worlds $(W, \le)$, described in Section~\ref{sec:knormas}) which appears indeed to be less expressive than the one available iALC. 
However, as opposed to iALC, IHL has support for (action-based) temporal reasoning. A characteristic that we believe makes it more suitable to model and reason about NorMAS than iALC.   

There is a line of thinking that agrees with the idea a norm should not have a truth-value, which we adopt in this paper.
%Although we mainly focus on one problem of the SDL (deontic paradoxes), such a logic presents other limitations. There are some lines of research, such as 
In~\cite{makinson1999,hansen2007,gabbay}, for instance, the authors advocate that SDL is not a suitable logic for modeling normative systems. In addition to the impossibility to represent CTD in a consistent way they agree that it makes no sense to assign norms truth values because norms are \emph{non-descriptive}, i.e., obligations and prohibitions (imperative norms) demand and allow behavior, respectively. They can be applied or not, can be followed or not and can be evaluated based on other norms (when a norm is judged based on a moral code, for instance). Some approaches in the literature have proposed to \emph{reconstruct} SDL as a logic of normative propositions (e.g.~\cite{makinson2000,aaqvist2008}) or as a logic of imperatives (e.g. ~\cite{hansen2006,hansen2008}) to cope with this problem, as pointed out in~\cite{gabbay}. We agree with their point of view that norms should not have truth value. However, we do not follow the proposal of reconstructing SDL. We chose a different approach, based on \emph{Intuitionistic Hybrid Logics} (IHL), where norms are nominals, as discussed in Section~\ref{sec:knormas}. 
%Second, as Deontic Logic is their chosen underlying logical framework, Deontic Logic paradoxes are inherited by NorMAS.

\section{Kelsenian Normative Multi-Agent Systems}\label{sec:knormas}

The Theory of Law, also called Jurisprudence, is interested in to determine the meaning of the concept of \emph{law}. Basically there are two views concerning this concept. The first view states that a law is a \emph{fact} that can be perceived by our senses. It is an object of natural sciences, such as, physics and chemistry. The second view states that a law is a norm, i.e., it is a rule that establishes what ought to be done. The Pure Theory of Law adopts the view that the law is a norm~\cite{kelsen1959}.

Kelsenian Jurisprudence~\cite{kelsen2005}, in a nutshell, advocates that ``the law'', in legal terms, or the norms in a NorMAS, is a set of individual regulatory statements, each of them created to enforce a positively desired behavior in the system. In~\cite{haeusler2010}, the authors propose the following requirements (or ontological commitments) from an analysis of Kelsenian Jurisprudence. Table~\ref{tab:kel-reg} interprets the three ontological commitments in~\cite{haeusler2010} in the context of NorMAS. Henceforth, we will refer to the contents of Table~\ref{tab:kel-reg} as \emph{Kelsenian regulation}.
\footnote{In this paper, we discuss Commitments (\ref{it:law-indiv}) and (\ref{it:law-prec}) only, without loosing consistency of presentation. Commitment (\ref{it:law-conn}) requires a broaden presentation on the interconnections of normative systems.}

\begin{table}[H]
\begin{enumerate}[(\bf I)]
\item\label{it:law-indiv} Individuals are norms;
\item\label{it:law-prec} There is a transitive and reflexive relationship between individuals and norms that reflects a precedence relationship between norms;
\item\label{it:law-conn} There are normative connections between individual norms in different normative systems or between different agent organizations in the same normative system.
\end{enumerate}
\caption{Kelsenian regulation}
\label{tab:kel-reg}
\end{table}

Ontological commitment (\ref{it:law-indiv}) is fulfilled by the choice of Hybrid Logic~\cite{prior1967,areces2001} (an extension to Modal Logics that adds a new sort of propositional symbols which are true at \emph{exactly} one possible world), and the representation of norms as nominals. In Hybrid Logics, there is a new kind of operator, called satisfaction operator $@_a$. It allows for the declaration of satisfaction statements $a : \varphi$ (sometimes written $@_a \varphi$), denoting that the formula $\varphi$ is true at the point to which the nominal $a$ refers to. More formally, the nominal $a$ represents a \emph{possible world} in the Kripke semantics of IHL. 

Before justifying the need for intuitionism, let us first recall the models of Classical Hybrid Logic. They are Kripke structures
$$(W, R, V)$$
where $W$ is a non-empty set of worlds, $R$ is a world accessibility relation $R \subseteq W \times W$ and $V : W \times \mathit{AP} \to \{0,1\}$, where $\mathit{AP}$ is the set of atomic propositions, together with a function $g$, called an assignment, that, to each nominal, assigns an element of $W$. An assignment $g'$ is an \emph{a-variant} of $g$ if $g'$ agrees with $g$ on all nominals save possibly $a$. The relation $M, g, w \models \varphi$ is defined by structural induction in the language of the Classical Hybrid Logic, where $M$ is a model, $g$ is an assignment, $w$ is an element of $W$, and $\varphi$ is a formula. The satisfaction relation for Classical Hybrid Logics is as follows: 

$$
\begin{array}{rcl}
M, g, w & \models & p \Leftrightarrow V(w, p)=1, \\
M, g, w & \models & a \Leftrightarrow w = g(a), \\
M, g, w & \models & \varphi \land \psi \Leftrightarrow M, g, w \models \varphi \mbox{ and } M, g, w \models \psi, \\
M, g, w & \models & \varphi \Rightarrow \psi \Leftrightarrow M, g, w \models \varphi \mbox{ implies } M, g, w \models \psi, \\
M, g, w & \models & \neg \varphi \Leftrightarrow M, g, w \not\models \varphi, \\
M, g, w & \models & \Box \varphi \Leftrightarrow  $for any element $v$ of  $W$ such that $wRv$, it is$ \\
 & & $ \hspace{20pt} the case that $ M, g, v  \models  \varphi, \\
M, g, w & \models & a:\varphi \Leftrightarrow M, g, g(a) \models \varphi. \\
\end{array}
$$

From the perspective of Kelsenian regulation, Classical Hybrid Logic is not enough because commitment (\ref{it:law-prec}) requires a pre-order among norms. Ordering is a natural way to prevent norm conflicts~\cite{santos2017} as it imposes a precedence between norms. Classical Hybrid Logic, however, does not offer it, as opposed to \emph{Intuitionistic} Hybrid Logic. 

%Heyting algebras~\cite{heyting1930}, models of Intuitionistic Logics~\cite{brauner2006}, provide such a structure: is a bounded lattice, with join and meet operations written $\sqcup$ and $\sqcap$, least element $\bot$, greatest element $\top$, and implication $\to$.

Formulas of Intuitionistic Hybrid Logic are the same as those of Classical Hybrid Logic.
However, connectives $\lor$ and $\diamondsuit$ are \emph{primitive} as they are not intuitionistically definable in terms of the other connectives, contrary to the classical case.

A model for Intuitionistic Hybrid Logic is a tuple 
$$(W, \le, \{D_w\}_{w\in W} ,\{\sim_w\}_{w\in W}, \{R_w\}_{w\in W}, \{V_w\}_{w\in W})$$ 
where $W$ is a non-empty set partially ordered by $\le$, for each $w$, $D_w$ is a non-empty set such that $w \le v$ implies $D_w \subseteq D_v$, for each $w$, $\sim_w$ is an equivalence relation on $D_w$ such that $w \le v$ implies $\sim_w \subseteq \sim_v$, for each $w$, $R_w$ is a binary relation on $D_w$ such that $w \le v$ implies $R_w \subseteq R_v$, and for each $w$, $V_w$ is a function that to each ordinary propositional symbol $p$ assigns a subset of $D_w$ such that $w \le v$ implies $V_w(p) \subseteq V_v(p)$.

$W$ represents the set of states of knowledge and for each state $w \in W$, $D_w$ denotes the set of possible worlds that are known in such a state. $V_w(p)$ denotes the set of worlds in which the proposition $p$ is known to be true.  

Given a model $$M = (W, \le, \{D_w\}_{w\in W} ,\{\sim_w\}_{w\in W}, \{R_w\}_{w\in W}, \{V_w\}_{w\in W})$$ and an element $w \in W$, a $w$-assignment is a
function $g$ that to each nominal assigns an element of $D_w$. Note that if $g$ is a $w$-assignment and $w \le v$, then $g$ is also a $v$-assignment (this is used in the clauses below for implication and the  $\Box$ operator). The relation $M, g, w,d \models \varphi$ is defined by induction, where $w$ is an element of $W$, $g$ is a $w$-assignment, $d$ is an element of $D_w$, and $\varphi$ is a formula.
%\begin{table}
$$
\begin{array}{rcl}
M, g, w, d & \models & p \Leftrightarrow d \in  V_w(p), \\
M, g, w, d & \models & a \Leftrightarrow d \sim_w g(a), \\
M, g, w, d & \models & \varphi \land \psi \Leftrightarrow M, g, w, d \models \varphi \mbox{ and } M, g, w, d \models \psi, \\
M, g, w, d & \models & \varphi \lor \psi \Leftrightarrow M, g, w, d \models \varphi \mbox{ or } M, g, w, d \models \psi, \\
M, g, w, d & \models & \varphi \Rightarrow \psi \Leftrightarrow \mbox{for all } v \ge w, M, g, v, d \models \varphi  $ implies $ \\ & & \hspace{48pt}  M, g, v, d \models \psi, \\
M, g, w, d & \models & \bot \Leftrightarrow \mbox{falsum} \\
M, g, w, d & \models & \Box \varphi \Leftrightarrow $ for any element $ v \ge w, $for all $ e \in D_v, \\ & & \hspace{27pt} d R_v e$ implies $M, g, v, e  \models  \varphi, \\
M, g, w, d & \models & \diamondsuit \varphi \Leftrightarrow \mbox{ for some $e \in D_w$, $d R_w e$ and $M, g, w, e \models \varphi$}, \\
M, g, w, d & \models & a:\varphi \Leftrightarrow M, g, w, g(a) \models \varphi. \\
\end{array}
$$

%Finally, commitment (\ref{it:law-conn}) requires a relation between named worlds. This is accomplished by the partial order $(W, \le)$ in the IHL model, with each $w \in W$ being a named world.
%One possibility to represent it would be by means of an assignment function that relates nominals with world elements. This aspect will be properly developed in future work.

We now define Kelsenian NorMAS in terms of an IHL model. (See Def.~\ref{def:knormas}.) Following Kelsen, for each norm we have a world $w \in W$ in the IHL model. 
%, ordered in $\le$ according to the order of declaration of the norms.
In other words, the pair $(W, \le)$ captures the normative part of the NorMAS. The MAS description is represented by the tuple $(D, R^{\mathit{Act}, A})$ of the IHL model, where $D$ denotes the set of the states of MAS and $R^{\mathit{Act}, A}$ its (action-labeled) transition relation, for actions in $\mathit{Act}$, defined as the disjoint union of the relations for each agent in $A$. (In this paper we consider relation $\sim$, in the IHL model, to be empty, without loss of generality, that is, there is no equivalence among states of the MAS.)

\begin{definition}[Kelsenian NorMAS]\label{def:knormas}
A Kelsenian Normative Multi-Agent System is an IHL model:

$$\mathcal{K} = (N, \le, D, \emptyset, \{R^{\mathit{Act}, A}_n\}_{n \in N}, V)$$ 

where $\mathit{Act}$ is a finite set of actions; and $A$ is a finite set of agents, such that, for all $n \in N$, $R^{\mathit{Act}, A}_n = \bigcup_{i = 1}^{\mid A \mid } R_i$, with $R_i \subseteq D \times \mathit{Act} \times D$ and $V : D \times \mathit{AP} \to \{0,1\}$ with $\mathit{AP}$ the set of atomic propositions.
\end{definition}

A state $n_i \in N$ is such that norm $n_i$ is \emph{upheld}. Note that upheld does not mean ``the norm holds'' as norms have no truth value. It means that $R_i$ complies with $n_i$, behaving accordingly to $n_i$.

\section{Kelsenian NorMAS for a CTD scenario}
\label{sec:CTDKnormas}

When we consider Normative MAS (NorMAS), scenarios where a norm regulated MAS may misbehave must be accounted for in a proper logical formalization. An example of misbehavior is when negative transitions are specified, assuming a labeled state-transition system formalization of MAS, and a conflict arises between MAS behavior and its regulation. In this section, we present an example that illustrates a situation in which an agent of a NorMAS chooses not to comply with a norm and there is a CTD norm saying what should be done if it occurs. We will consider in this example the Contract Net~\cite{contract-net} agent interaction protocol, standardized by the Foundation for Intelligent Physical Agents (FIPA).

\begin{figure}[htp]
\begin{adjustbox}{max totalsize={.55\textwidth}{\textheight},center}
\begin{tikzpicture}[->,>=stealth',shorten >=2pt,auto,node distance=4cm, semithick]
\node[state] (I) {$I$} ;
\node[state] (P1) [right of = I] {$P_1$} ;
\node[state] (P2) [below of = P1] {$P_2$} ;
\node at ([shift={(270:1.0)}] P2.270) {$\ldots$} ;
\node[state] (Pm) [below of = P2] {$P_m$} ;
\path (I) edge [loop left] node [left] {\emph{recognizes}} (I) ;
{%\color{blue}
\path [dashed] (I) edge node {\emph{announce}} (P1) ;
\path [dashed] (I) edge node [right] {\emph{announce}} (P2) ;
\path [dashed] (I) edge node [below] {\emph{announce}} (Pm) ;}
{%\color{green}
\path [->>] (P1) edge [bend right] node [above] {\emph{bid}} (I) ;
\path [->>] (P2) edge [bend right] node [above] {\emph{bid}} (I) ;
\path [->>] (Pm) edge [bend left] node [below] {\emph{bid}} (I) ;}
{%\color{red}
\path (I) edge [bend left=90] node [above] {\emph{award}($P_1$)} (P1) ;
\path (I) edge [bend right] node [right] {\emph{award}($P_1$)} (P2) ;
\path (I) edge [bend right=90] node [below] {\emph{award}($P_1$)} (Pm) ;}
\path (P1) edge [loop right] node [right] {\emph{recognizes}} (P1) ;
\end{tikzpicture}
\end{adjustbox}
\caption{An instance of the Contract Net protocol}\label{fig:contract-net}
\end{figure}
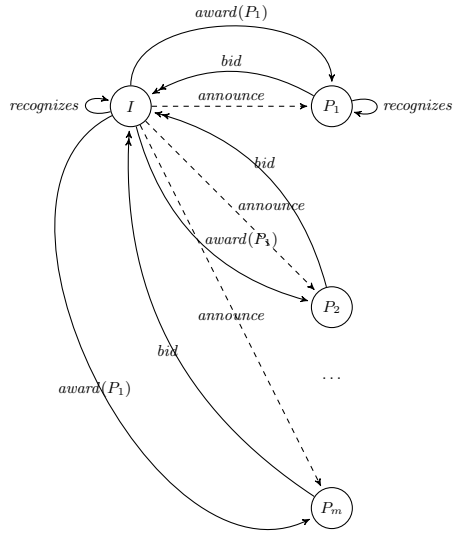

 Informally, the Contract Net protocol specifies that: (i) when an (initiator) agent \emph{realizes} it has a problem to solve, (ii) it may \emph{announce} it to other agents, (iii) which in turn will \emph{bid} for the task of solving (part) of the given problem, (iv) being then notified by the initiator which was the \emph{awarded} agent, that then (v) \emph{expedites} solving the problem. Figure~\ref{fig:contract-net} illustrates the protocol. 

Let us consider now a subclass of Contract Net implementations regulated by the norms in Table~\ref{tab:contract-net-norms}.
\begin{table}[H]
\begin{center}
\begin{tabular}{c | l}
Norm\#& Norm description \\ \hline
$n_1$. & ``Once a problem is announced then all agents must \\& \quad  bid.'' \\
$n_2$. & ``Once a problem is announced, all agents must bid \\ & \quad and then an agent must be awarded.'' \\
$n_3$. & ``If it is not the case that a problem was announced\\ & \quad and agents bade for it then there must not be an \\ & \quad awarded agent.''
\end{tabular}
\end{center}
\caption{A regulated Contract Net}
\label{tab:contract-net-norms}
\end{table}
Now, if in an implementation of Contract Net with norms from Table~\ref{tab:contract-net-norms}, say an electronic commerce system, an agent refuses to perform action \emph{bid}, it would give rise to a misbehaving NorMAS, as action \emph{bid}, at the same time, must and does not occur. This scenario is depicted in Figure~\ref{fig:conflict} where $P_3$ is the misbehaving agent that does not implement action \emph{bid}, with the interrupted or negative transition (loosely dashed, with double tip and a ray symbol denoting interruption) denoting that the agent did not perform action \emph{bid}. 
%(Note that specifying a negative transition is part of the \emph{description} of what the system does (or does not), not its \emph{prescription}, or what properties it should have.)
\usetikzlibrary{shapes.geometric,arrows.meta,decorations.markings}
\tikzset{
  big dot/.style={circle, draw, inner sep=0pt, minimum size=3mm, fill=yellow},
  input/.style={draw, trapezium, trapezium left angle=60, trapezium right angle=120,
    line width=1.2pt, fill={rgb:black,1;white,2}},
  obstacle/.style={draw, trapezium, trapezium left angle=120, trapezium right angle=60,
    line width=1.2pt, fill={rgb:black,1;white,2}},
  flash/.style args={#1:#2}{postaction=decorate,decoration={name=markings,
    mark=at position #1 with {%
    \draw[fill=#2, line width=.75\pgflinewidth, line cap=round, line join=round]
         (+\pgflinewidth,+7\pgflinewidth)   -- ++ ( left:+2\pgflinewidth) 
      -- (+-4\pgflinewidth,+-\pgflinewidth) -- ++ (right:+5\pgflinewidth)
      -- (+-\pgflinewidth,+-7\pgflinewidth) -- ++ (right:+2\pgflinewidth)
      -- (+4\pgflinewidth,\pgflinewidth)    -- ++ (left:+5\pgflinewidth)
      -- cycle;}}}
}
\begin{figure}[ht]
\begin{adjustbox}{max totalsize={.55\textwidth}{\textheight},center}
\begin{tikzpicture}[->,>=stealth',shorten >=2pt,auto,node distance=4cm, semithick]
\node[state] (I) {$I$} ;
\node[state] (P3) [left of = I] {$P_3$} ;
\node[state] (P1) [right of = I] {$P_1$} ;
\node[state] (P2) [below of = P1] {$P_2$} ;
\node at ([shift={(270:1.0)}] P2.270) {$\ldots$} ;
\node[state] (Pm) [below of = P2] {$P_m$} ;
\path [->>] (P3) edge [bend left=90] node [above] {\emph{bid}} (I) ;
{%\color{blue}
\path [dashed] (I) edge node {\emph{announce}} (P1) ;
\path [dashed] (I) edge node [right] {\emph{announce}} (P2) ;
\path [dashed] (I) edge node [above] {\emph{announce}} (P3) ;
\path [dashed] (I) edge node [below] {\emph{announce}} (Pm) ;}
{%\color{green}
\path [->>] (P1) edge [bend right] node [above] {\emph{bid}} (I) ;
\path [->>] (P2) edge [bend right] node [above] {\emph{bid}} (I) ;
\path [loosely dashed, ->>] (P3) edge [bend left, flash=.5:black] node [above] {\emph{bid}} (I) ;
\path [->>] (Pm) edge [bend left] node [below] {\emph{bid}} (I) ;}
{%\color{red}
\path (I) edge [bend left=90] node [above] {\emph{award}($P_1$)} (P1) ;
\path (I) edge [bend right] node [right] {\emph{award}($P_1$)} (P2) ;
\path (I) edge [bend right=90] node [below] {\emph{award}($P_1$)} (Pm) ;}
\end{tikzpicture}
\end{adjustbox}
\caption{A misbehaving Contract Net implementation}\label{fig:conflict}
\end{figure}
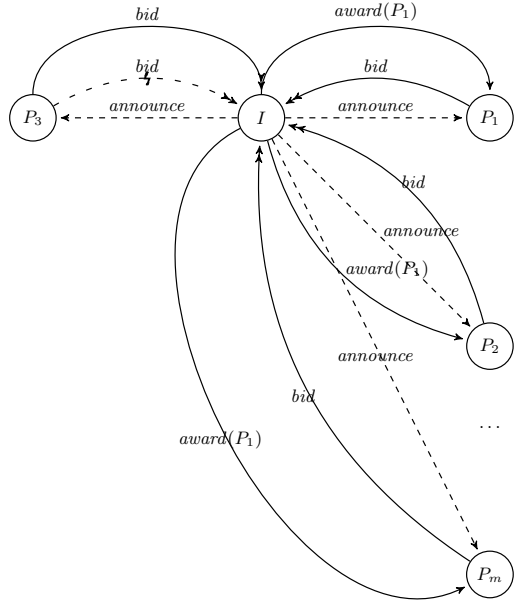

What we have just described is an example of the CTD scenarios that become paradoxical when Deontic Logic is chosen as the underlying logic to model and reason on NorMAS. More precisely, to see that the misbehaved Contract Net implementation in Figure~\ref{fig:conflict} is an instance of the so-called Chisholm paradox (see Section~\ref{sec:intro}) we just need to replace the proposition $p$ from the Chisholm paradox  for ``Once a problem is announced then all agents bid'', and proposition $q$ for ``an agent is awarded''. When action \emph{bid} is not performed by an agent, thus leading to the negative transition in Figure~\ref{fig:conflict}, ``all agents must bid'' becomes false and so does $p$. 

\begin{figure}[H]
\begin{adjustbox}{max totalsize={.55\textwidth}{\textheight},center}
\begin{tikzpicture}[->,>=stealth',shorten >=2pt,auto,node distance=4cm, semithick]
\node[elliptic state] (n1) {$n_1 \models \top$} ;
\node[elliptic state] (n2) [right of = n1] {$n_2 \models \top$} ;
\node[elliptic state] (n1Mn2) [below right of = n1] {$n_1 \sqcap n_2 \models \top$} ;
\node[elliptic state] (n3) [right of = n1Mn2] {$n_3\models \neg p$} ;
\node[elliptic state] (n1Mn2Mn3) [below right of = n1Mn2] {$n_1 \sqcap n_2 \sqcap n_3 \not\models p$} ;
\path (n1Mn2Mn3) edge node {$\le$} (n1Mn2) ;
\path (n1Mn2Mn3) edge node [right] {$\le$} (n3) ;
\path (n1Mn2) edge node [right] {$\le$} (n1) edge node [right] {$\le$} (n2) ;
\end{tikzpicture}
\end{adjustbox}
\caption{A Kelsenian NorMAS for a normative contradiction}
\label{fig:regulated-contract-net-model}
\end{figure}
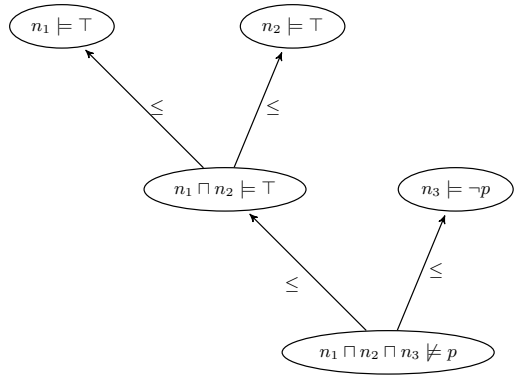

\begin{example}[Kelsenian NorMAS for the misbehaving Contract Net]
The Kelsenian NorMAS for the Contract Net instance in Figure~\ref{fig:conflict} is the tuple $$\mathcal{K} = (N, \le, D, \emptyset, \{R^{\mathit{Act}, A}_n\}_{n \in N}, V)$$ where $N = \{n_1, n_2, n_3,$  $n_1 \sqcap n_2, n_1 \sqcap n_2 \sqcap n_3\}$, where operation $\sqcap$ denotes the \emph{meet} operation of IHL's underlying Heyting algebra~\cite{heyting1930}.
(Kripke models for Intuitionistic Logics are Heyting algebras, and therefore lattices.) A state $n_i$ is such that norm $n_i$ is upheld. 
%Note that upheld does not mean ``the norm holds'' as norms have no truth value. It means that $R$ is $n_i$ is such that agents comply with $n_i$. 
The ordering $\le$ is as pictured in Figure~\ref{fig:regulated-contract-net-model}, $D = \{\mathit{Recognized}, \mathit{Announced}, \mathit{Bade}, \mathit{Awarded}\}$, relation $R$ is pictured in Figure~\ref{fig:conflict}, $A = \{I, P_1, P_2, \ldots, P_n\}$, $\mathit{Act} = \{\mathit{recognize}, \mathit{announce}, \mathit{bid}, \mathit{award}\}$, and $V = \emptyset$. Let $p$ denote ``Once a problem is announced then all agents bid''.

Figure~\ref{fig:regulated-contract-net-model} pictures an IHL model for a NorMAS that is an instance of the Chisholm paradox. In this model each norm is represented by a nominal (world), then $n_1 \models \top$ and $n_2 \models \top$ indicate the existence of norms $n_1$ and $n_2$ in the NorMAS. The meet of these two worlds $n_1 \sqcap n_2 \models \top$ is a world in which it is obligatory to comply with $n_1$ and with $n_2$. The nominal $n_3$ is the world in which $\neg p$ holds (and then proposition $p$ does not hold), that is, it represents a case where the agency does not comply with the first norm of Table~\ref{tab:contract-net-norms}.\footnote{Note that not being a model for $p$ and $\neg p$ holding in a state are different things. In the former case state $n_1 \sqcap n_2 \sqcap n_3$ is \emph{not} a model for $p$ while in the latter case state $n_3$ \emph{is} a model for $\neg p$, that is, $\neg p$ is satisfiable.}

The least element of $(N, \le)$ is $n_1 \sqcap n_2 \sqcap n_3$, that is, the meet of $n_1$, $n_2$ and $n_3$ denoting a world which is not a model for ``Once a problem is announced then all agents bid'' since it is ensured that $\neg p$ holds.  

\end{example}

\section{Hill's taxonomy and Kelsenian NorMAS}\label{sec:hill-knormas}

In~\cite{hill} H. Hamner Hill proposes a taxonomy for normative conflict, an important reference in the normative systems literature in the context of norm conflict identification and resolution~\cite{santos2017}. Hill argues, developing ideas by Kelsen and others, that normative conflicts identified by impossibility-of-joint-compliance test, when it is impossible for a norm subject to comply with both of a pair of norms, is too restrictive. Let us briefly recall why.

The impossibility-of-joint-compliance test can only be applied to norms that one can construct obedience statements for, that is, a statement that certifies compliance of a norm subject with a given norm. For example, let us consider norm $n_1$ from Table~\ref{tab:contract-net-norms} from Section~\ref{sec:CTDKnormas}, that says ``Once a problem is announced then all agents must bid''. Its obedience statement would be ``A problem has been announced and all agents bade''. Therefore, impossibility-of-joint-compliance test only applies to deontic imperatives, i.e., it cannot detect conflicts involving permissions. 

There are, however, scenarios where: (i) deontic permissions conflict with deontic permissions, (ii) there are regulatory modalities other than the deontic modalities, as in power-conferring norms~\cite{hart}, and (iii) conflicts considering non-deontic norms.

Quoting Hill~\cite[pg. 238 and 239]{hill}:
\begin{quote}
For Kelsen, a normative conflict is a clash of forces, forces which operate in different directions in a single point. ($\ldots$) The generic phenomenon of normative conflict occurs when norms interact in ways that the function of one or more of the norms involved is thwarted.  
\end{quote} 

In order to be able to detect conflicts involving permissions, Hill replaced the concept of obedience by the concept of conformity. For each norm that indicates permission, there are two conformity statements, as follows: (i) the first conformity statement indicates that the agent chose to do what is being allowed by the permission; and (ii) the second conformity statement indicates that the agent chose not to do what is being allowed by the permission. Then, considering a scenario in which there is a prohibition stating that \textit{a} cannot be done and there is a permission stating that \textit{a} can be done. In this example, there is a conflict between the prohibition and the first conformity statement of the permission.

Hill defines a functional taxonomy of normative conflicts: (i) normative contradiction, (ii) normative collision, and (iii) normative competition. Normative contradictions are ``a purely deontic phenomena'' where only deontic norms may contradict each another. In normative collisions, deontic imperatives and deontic permissions collide, mixtures of deontic and non-deontic norms collide or only non-deontic norms collide. Normative competition regards norm conflicts from distinct normative systems. In what follows, we discuss about Kelsenian NorMAS for normative contradictions and normative collisions between deontic imperatives and deontic permissions. In this paper we do not consider normative competitions since we are only interested in dealing with conflicts in one NorMAS.  

\paragraph{Normative contradiction.} Normative contradictions are conflicts that can be detected by the impossibility-of-joint-compliance test. They come from scenarios where the norm subject finds oneself faced with duties one cannot fulfil. Normative contradictions violate the principle of consistency of SDL that says:  $O (a) \Rightarrow \neg O (\neg a)$. This kind of conflict usually occurs between norms imposed by different authorities. In SDL, normative contradiction is a conflict in the form:

$$O (a) \land O (\neg a)$$

Note that, the so-called contrary-to-duty paradoxes, when SDL is used to formalize them, result in this kind of conflict. One such contrary-to-duty paradox is Chisholm paradox which is embodied in the ``misbehaved'' Contract Net implementation in Figure~\ref{fig:conflict}. When formalized in Deontic Logic, it gives rise to an inconsistent theory. An example of Kelsenian NorMAS model for a normative contradiction is illustrated in Figure~\ref{fig:regulated-contract-net-model}.

\paragraph{Normative collision.} 
Normative collisions are conflicts that do not involve logic inconsistency but involve functional incompatibility. In SDL, they are conflicts in the form:

$$O (\neg a) \land P (a)$$ or similarly,

$$O (a) \land P (\neg a)$$

where \textit{P} denotes the deontic modality for permission.

Following the same approach for deontic imperatives, in our approach deontic permissions do not have truth value: there are nominal states denoting when the agency avails itself of a permission and otherwise. 

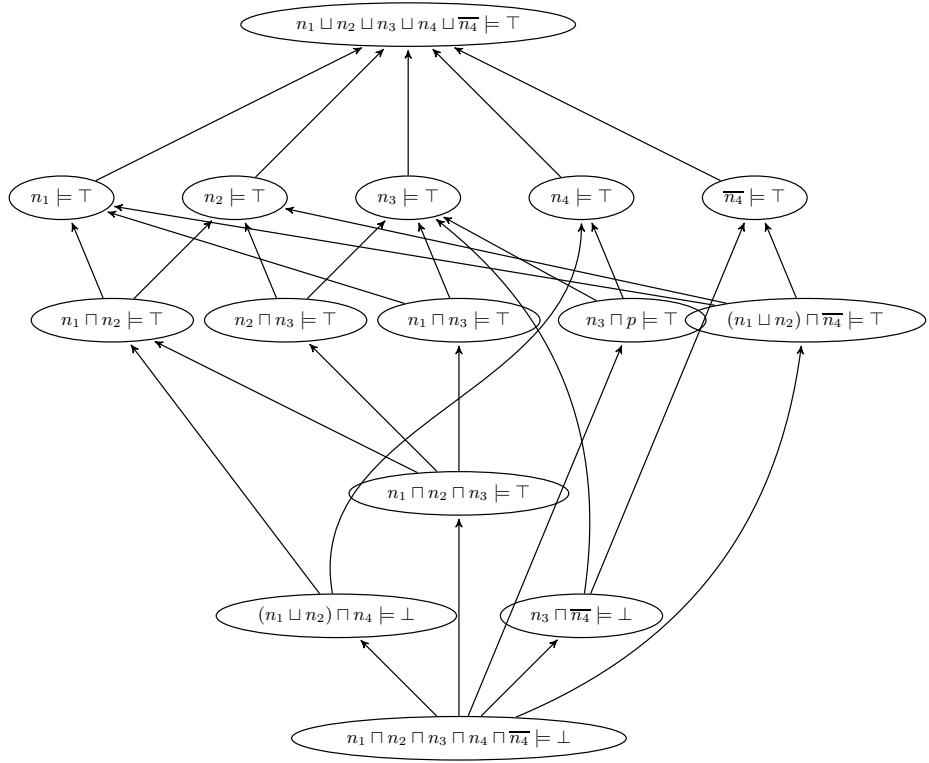
\begin{figure}[ht]
\begin{adjustbox}{max totalsize={\textwidth}{\textheight},left}
\begin{tikzpicture}[->,>=stealth',shorten >=1.5pt, auto, node distance=3.0cm, semithick]
\node[elliptic state] (n1Jn2Jn3JpJpb) {$n_1 \sqcup n_2 \sqcup n_3 \sqcup n_4 \sqcup \overline{n_4} \models \top$} ;
\node[elliptic state] (n3) [below of = n1Jn2Jn3JpJpb] {$n_3 \models \top$} ;
\node[elliptic state] (p) [right of = n3] {$n_4 \models \top$} ;
\node[elliptic state] (n2) [left of = n3] {$n_2 \models \top$} ;
\node[elliptic state] (n1) [left of = n2] {$n_1 \models \top$} ;
\node[elliptic state] (pb) [right of = p] {$\overline{n_4} \models \top$} ;
\node[elliptic state] (n2Mn3) [below left of = n3] {$n_2 \sqcap n_3 \models \top$} ;
\node[elliptic state] (n1Mn2) [left of = n2Mn3] {$n_1 \sqcap n_2 \models \top$} ;
\node[elliptic state] (n1Mn3) [right of = n2Mn3] {$n_1 \sqcap n_3 \models \top$} ;
\node[elliptic state] (n1Mn2Mn3) [below of = n1Mn3] {$n_1 \sqcap n_2 \sqcap n_3 \models \top$} ;
\node[elliptic state] (n3Mp) [right of = n1Mn3] {$n_3 \sqcap p \models \top$} ;
\node[elliptic state] (n1Jn2Mpb) [right of = n3Mp] {$(n_1 \sqcup n_2) \sqcap \overline{n_4} \models \top$} ;
\node[elliptic state] (n1Jn2Mp) [below left of = n1Mn2Mn3] {$(n_1 \sqcup n_2) \sqcap n_4 \models \bot$} ;
\node[elliptic state] (n3Mpb) [below right of = n1Mn2Mn3] {$n_3 \sqcap \overline{n_4} \models \bot$} ;
\node[elliptic state] (n1Mn2Mn3MpMpb) [below right of = n1Jn2Mp] {$n_1 \sqcap n_2 \sqcap n_3 \sqcap n_4 \sqcap \overline{n_4} \models \bot$} ;
\path (n1) edge  (n1Jn2Jn3JpJpb) ;
\path (n2) edge   (n1Jn2Jn3JpJpb) ;
\path (n3) edge   (n1Jn2Jn3JpJpb) ;
\path (p) edge   (n1Jn2Jn3JpJpb) ;
\path (pb) edge   (n1Jn2Jn3JpJpb) ;
\path (n1Mn2) edge   (n1) edge   (n2) ;
\path (n1Mn3) edge   (n1) edge   (n3) ;
\path (n2Mn3) edge   (n2) edge   (n3) ;
\path (n3Mp) edge   (n3) edge   (p) ;
\path (n1Jn2Mpb) edge   (n1) edge   (n2)  edge (pb);
\path (n1Mn2Mn3) edge   (n1Mn2) edge   (n2Mn3)  edge (n1Mn3);
\path (n1Jn2Mp) edge (n1Mn2) edge [in=270, out=100] (p) ;
\path (n3Mpb) edge [bend right] (n3) edge (pb) ;
\path (n1Mn2Mn3MpMpb) edge   (n1Jn2Mp) edge   (n3Mpb) edge (n1Mn2Mn3) edge (n3Mp) edge [bend right] (n1Jn2Mpb);
%\path (n1Mn2Mp2) edge node [right] {$\le$} (n1Mp2) edge node [left] {$\le$} (n2Mp2) ;
%\path (n2Mn3Mp2) edge node [right] {$\le$} (n2Mp2) edge node [left] {$\le$} (n3Mp2) ;
%\path (n1Mp2) edge node [right] {$\le$} (p2) ;
%\path (n2Mp2) edge node [right] {$\le$} (p2) ;
%\path (n3Mp2) edge node [left] {$\le$} (p2) ;
\end{tikzpicture}
\end{adjustbox}
\caption{A Kelsenian NorMAS for a normative collision}
\label{fig:normative-collision}
\end{figure}

We can illustrate this situation by considering Contract Net instance from Figure~\ref{fig:contract-net}, the norms from Table~\ref{tab:contract-net-norms}, and a permission \emph{prescribing} that ``An agent is allowed not to bid". Note that this normative system is slightly different from the ``misbehaving'' Contract Net instance in Figure~\ref{fig:conflict}. There, the negative transition is part of the \emph{description} of the system (the behavior of the system) and here the permission not to act is at the \emph{prescription} (what it should do) level. There is clearly a normative collision here between norm $n_1$ and the given permission, if the agent choose to do what is being permitted (not to bid). Figure~\ref{fig:regulated-contract-net-model} pictures the IHL model for the normative collision example where world $n_i$ denotes norm $n_i$, with $i \in \{1,2,3\}$, $n_4$ denotes the state where the agency avails itself of the given permission and $\overline{n_4}$ otherwise. The resulting Kelsenian NorMAS is a lattice with combinations of $n_i$, $n_4$, and $\overline{n_4}$, as depicted in Figure~\ref{fig:normative-collision}.  (All arrows denote relation $\le$.)

The infimum of the lattice is $n_1 \sqcap n_2 \sqcap n_3 \sqcap n_4 \sqcap \overline{n_4} \models \bot$ denoting a state of the Kelsenian NorMAS where no property holds since a join between states $n_4$ and $\overline{n_4}$ results in a state where agents  chose not to bid and chose to bid making it impossible to uphold norms $n_1$ to $n_3$. In particular, either norm $n_1$ or $n_2$ can not be upheld when agents choose not to bid ($n_4$) as they must when a problem is announced. (A similar argument is used to justify state $n_3 \sqcap \overline{n_4} \models \bot$.) On the other hand, when either $n_1$ or $n_2$ can be upheld in a state when agents may bid and therefore $(n_1 \sqcup n_2) \sqcap \overline{n_4} \models \top$. Norms are upheld in states where single norms  alone are upheld (not a join or meet of worlds) or a permission is upheld, such as $n_1$ or $n_4$. The join of all the worlds (not to be confused with their conjunction) represents the supremum of the lattice.  

\section{Conclusion}\label{sec:conclusion}

Regulatory mechanisms are required in order to deal with the autonomy and (possible) heterogeneity of software agents that compose a Multi-Agent System (MAS). Norms can be applied in such systems as way of coordination. They are social constraints that establish explicitly, for instance, which actions a software agent is permitted, prohibited or obliged to perform.

Standard techniques for modeling and reasoning on Normative Multi-Agent Systems (NorMAS) fall prey of the interpretation that norms are formulas. When Deontic Logic is chosen as underlying formalism for NorMAS, CTD scenarios cannot be represented because its SDL formalization is inconsistent. This is a relevant problem in NorMas, where autonomous software agents should be able to violate norms. In order to cope with this problem, the approach presented in this paper adopts Intuitionistic Hybrid Logic as underlying logic, meeting the commitments of Kelsenian Jurisprudence. 

An important issue in systems regulated by multiple norms, is the possibility of conflicts among norms. We discuss the possible kinds of conflicts, according to the Hill's taxonomy, that may arise in a system governed by multiple norms and show how the approach presented allows us to construct models for conflicts involving deontic imperatives and deontic permissions in an elegant and simple way. In order to exemplify our IHL model, we use a Contract Net scenario that is a FIPA protocol for multi-agent communication.

Future work includes further developing our framework, in particular regarding the normative connections from Commitment~\ref{it:law-conn} possibly in the directions of iALC where relations between concepts denote classes of norms. Another direction is the automation of our approach to simulate and (bound) model check Kelsenian NorMAS where the model is the semantics of a description in an appropriate specification language.

%%%%%%%%%%%%%%%%%%%%%%%%%%%%%%%%%%%%%%%%%%%%%%%%%%%%%%%%%%%%%%%%%%%%%%%%%%%%%%%%%%%%%%%%%%%%%%%%%%%%%%%%%
%% start of main body of paper

%\input{samplebody-conf}

%%%%%%%%%%%%%%%%%%%%%%%%%%%%%%%%%%%%%%%%%%%%%%%%%%%%%%%%%%%%%%%%%%%%%%%%%%%%%%%%%%%%%%%%%%%%%%%%%%%%%%%%%
%% bibliography: see CFP for number of permitted pages

%\bibliographystyle{ACM-Reference-Format}  % do not change this line!
%\bibliographystyle{abbrv}
%\bibliography{aamas-2018-bibliography}  % put name of your .bib file here

\begin{thebibliography}{10}

\bibitem{aaqvist2008}
L.~{\AA}qvist.
\newblock Alchourr{\'o}n and bulygin on deontic logic and the logic of
  norm-propositions: Axiomatization and representabeility results.
\newblock {\em Logique et Analyse}, 51(203):225--261, 2008.

\bibitem{areces2001}
C.~Areces, P.~Blackburn, and M.~Marx.
\newblock Hybrid logics: Characterization, interpolation and complexity.
\newblock {\em The Journal of Symbolic Logic}, 66(3):977--1010, 2001.

\bibitem{brauner2006}
T.~Bra{\"u}ner and V.~de~Paiva.
\newblock Intuitionistic hybrid logic.
\newblock {\em Journal of Applied Logic}, 4(3):231--255, 2006.

\bibitem{broersen2013}
J.~Broersen, S.~Cranefield, Y.~Elrakaiby, D.~Gabbay, D.~Grossi, E.~Lorini,
  X.~Parent, L.~W.~N. van~der Torre, L.~Tummolini, P.~Turrini, and
  F.~Schwarzentruber.
\newblock {Normative Reasoning and Consequence}.
\newblock In G.~Andrighetto, G.~Governatori, P.~Noriega, and L.~W.~N. van~der
  Torre, editors, {\em Normative Multi-Agent Systems}, volume~4 of {\em
  Dagstuhl Follow-Ups}, pages 33--70. Schloss Dagstuhl--Leibniz-Zentrum fuer
  Informatik, Dagstuhl, Germany, 2013.

\bibitem{gabbay}
J.~Broersen, S.~Cranefield, Y.~Elrakaiby, D.~Gabbay, D.~Grossi, E.~Lorini,
  X.~Parent, L.~W.~N. van~der Torre, L.~Tummolini, P.~Turrini, and
  F.~Schwarzentruber.
\newblock {Normative Reasoning and Consequence}.
\newblock In G.~Andrighetto, G.~Governatori, P.~Noriega, and L.~W.~N. van~der
  Torre, editors, {\em Normative Multi-Agent Systems}, volume~4 of {\em
  Dagstuhl Follow-Ups}, pages 33--70. Schloss Dagstuhl--Leibniz-Zentrum fuer
  Informatik, Dagstuhl, Germany, 2013.

\bibitem{carmo2002}
J.~Carmo and A.~J. Jones.
\newblock Deontic logic and contrary-to-duties.
\newblock {\em Handbook of philosophical logic}, 8:265--343, 2002.

\bibitem{castro2009deontic}
P.~F. Castro and T.~S.~E. Maibaum.
\newblock Deontic logic, contrary to duty reasoning and fault tolerance.
\newblock {\em Electronic Notes in Theoretical Computer Science},
  258(2):17--34, 2009.

\bibitem{castro2009}
P.~F. Castro and T.~S.~E. Maibaum.
\newblock Reasoning about system-degradation and fault-recovery with deontic
  logic.
\newblock In M.~J. Butler, C.~B. Jones, A.~Romanovsky, and E.~Troubitsyna,
  editors, {\em Methods, Models and Tools for Fault Tolerance}, volume 5454 of
  {\em Lecture Notes in Computer Science}, pages 25--43. Springer, Berlin,
  Heidelberg, 2009.

\bibitem{chisholm1963}
R.~M. Chisholm.
\newblock Contrary-to-duty imperatives and deontic logic.
\newblock {\em Analysis}, 24(2):33--36, 1963.

\bibitem{cholvy1995}
L.~Cholvy and F.~Cuppens.
\newblock Solving normative conflicts by merging roles.
\newblock In {\em Proceedings of the 5th International Conference on Artificial
  Intelligence and Law}, ICAIL '95, pages 201--209, New York, NY, USA, 1995.
  ACM.

\bibitem{contract-net}
FIPA.
\newblock Fipa contract net interaction protocol specification.
\newblock {\em Technical report, Foundation for Intelligent Physical Agents},
  2002.

\bibitem{haeusler2010}
E.~H. Haeusler, V.~de~Paiva, and A.~Rademaker.
\newblock Using intuitionistic logic as a basis for legal ontologies.
\newblock In {\em Proceedings of the 4th Workshop on Legal Ontologies and
  Artificial Intelligence Techniques}, pages 69--76, Fiesole (Florence), Italy,
  2010. European University Institute.

\bibitem{haeusler2011}
E.~H. Haeusler, V.~de~Paiva, and A.~Rademaker.
\newblock Intuitionistic description logic and legal reasoning.
\newblock In {\em Database and Expert Systems Applications (DEXA), 2011 22nd
  International Workshop on}, pages 345--349, Toulouse, France, 2011. IEEE,
  IEEE Computer Society.

\bibitem{hansen2006}
J.~Hansen.
\newblock Deontic logics for prioritized imperatives.
\newblock {\em Artificial Intelligence and Law}, 14(1):1--34, 2006.

\bibitem{hansen2008}
J.~Hansen.
\newblock Prioritized conditional imperatives: problems and a new proposal.
\newblock {\em Autonomous Agents and Multi-Agent Systems}, 17(1):11--35, 2008.

\bibitem{hansen2007}
J.~Hansen, G.~Pigozzi, and L.~van~der Torre.
\newblock Ten philosophical problems in deontic logic.
\newblock In G.~Boella, L.~van~der Torre, and H.~Verhagen, editors, {\em
  Normative Multi-agent Systems}, number 07122 in Dagstuhl Seminar Proceedings,
  Dagstuhl, Germany, 2007. Internationales Begegnungs- und Forschungszentrum
  f{\"u}r Informatik (IBFI), Schloss Dagstuhl.

\bibitem{hart}
H.~L.~A. Hart and L.~Green.
\newblock {\em The concept of law}.
\newblock Oxford University Press, 2012.

\bibitem{heyting1930}
A.~Heyting.
\newblock Die formalen regeln der intuitionistischen logik.
\newblock {\em Sitzungsberichte der Preussischen Akademie der Wissenshaften,
  physikalisch-mathematische Klasse}, 42:158--169, 1930.

\bibitem{hill}
H.~H. Hill.
\newblock A functional taxonomy of normative conflict.
\newblock {\em Law and Philosophy}, 6(2):227--247, 1987.

\bibitem{kelsen1959}
H.~Kelsen.
\newblock What is the pure theory of law.
\newblock {\em Tul. L. Rev.}, 34:269, 1959.

\bibitem{kelsen2005}
H.~Kelsen.
\newblock Pure theory of law, trad. de la 2. {\textordfeminine} ed.
\newblock {\em M. Knight. NJ: The Law Book Exchange}, 192, 2005.

\bibitem{makinson1999}
D.~Makinson.
\newblock On a fundamental problem of deontic logic.
\newblock {\em Norms, Logics and Information Systems. New Studies on Deontic
  Logic and Computer Science}, pages 29--54, 1999.

\bibitem{makinson2000}
D.~Makinson and L.~Van Der~Torre.
\newblock Input/output logics.
\newblock {\em Journal of Philosophical Logic}, 29(4):383--408, 2000.

\bibitem{mcnamara2006}
P.~McNamara.
\newblock Deontic logic.
\newblock {\em Handbook of the History of Logic}, 7:197--288, 2006.

\bibitem{MeyerDignumWieringa1994}
J.-J. Meyer, F.~Dignum, and R.~Wieringa.
\newblock {\em The paradoxes of deontic logic revisited: a computer science
  perspective}.
\newblock Number UU-CS-1994-38 in Technical Report. University of Utrecht, 9
  1994.

\bibitem{prior1967}
A.~N. Prior.
\newblock {\em Past, present and future}, volume 154.
\newblock Clarendon Press Oxford, 1967.

\bibitem{santos2017}
J.~S. Santos, J.~O. Zahn, E.~A. Silvestre, V.~T. Silva, and W.~W. Vasconcelos.
\newblock Detection and resolution of normative conflicts in multi-agent
  systems: a literature survey.
\newblock {\em Autonomous Agents and Multi-Agent Systems}, 31:1236--1282, 2017.

\bibitem{vasconcelos2009}
W.~W. Vasconcelos, M.~J. Kollingbaum, and T.~J. Norman.
\newblock Normative conflict resolution in multi-agent systems.
\newblock {\em Autonomous agents and multi-agent systems}, 19(2):124--152,
  2009.

\bibitem{vonWright1951}
G.~H. von Wright.
\newblock Deontic logic.
\newblock {\em Mind}, 60(237):1--15, 1951.

\end{thebibliography}

\end{document}